\documentclass[twocolumn,english,showpacs,footinbib,aps,prb]{revtex4-1}

\usepackage[normalem]{ulem}
 
\usepackage[T1]{fontenc}
\usepackage[latin9]{inputenc}
\usepackage{graphicx}
\usepackage{amsmath}
\usepackage{amssymb}
\usepackage{bbm}
\usepackage{xspace}
\usepackage{color}
\usepackage{footnote}
\usepackage{comment}
\usepackage{hyperref}
\usepackage{subfigure}

\definecolor{dgreen}{rgb}{0.0,0.5,0.0}

\definecolor{orange}{RGB}{252,77,6}
\definecolor{brown}{RGB}{200,127,50}
\definecolor{blue}{RGB}{00,000,100}
\definecolor{blue2}{RGB}{00,000,250}
\definecolor{green1}{RGB}{00,100,00}
\definecolor{green2}{RGB}{00,150,00}
\definecolor{green3}{RGB}{00,200,00}
\definecolor{green4}{RGB}{00,250,00}

\newcommand{\Fig}[1]{Fig.\thinspace{}\ref{#1}}
\newcommand{\eq}[1]{Eq.\thinspace{}(\ref{#1})}

\newcommand{\se}{Sec.\@\xspace}

\newcommand{\etal}[0]{\textit{et al.}}

\newcommand{\beq}{\begin{equation}}
\newcommand{\eeq}{\end{equation}}

\newcommand{\Imm}{{\rm Im\:}}

\newcommand{\tr}[1]
{
\text{tr}\,#1 
}
\newcommand{\nag}{{\phantom{\dag}}}

\begin{document}

\title{Transport Through a Correlated Interface:  Auxiliary Master Equation Approach }

\author{Irakli Titvinidze}

\email{irakli.titvinidze@tugraz.at}


\affiliation{Institute of Theoretical and Computational Physics, Graz University
of Technology, 8010 Graz, Austria}

\author{Antonius Dorda}

\affiliation{Institute of Theoretical and Computational Physics, Graz University
of Technology, 8010 Graz, Austria}

\author{Wolfgang von der Linden}

\affiliation{Institute of Theoretical and Computational Physics, Graz University
of Technology, 8010 Graz, Austria}

\author{Enrico Arrigoni}

\affiliation{Institute of Theoretical and Computational Physics, Graz University
of Technology, 8010 Graz, Austria}


\begin{abstract} 
We present  {improvements} of a recently introduced numerical method  [{\color{blue}Arrigoni \etal{}, Phys. Rev. Lett. \textbf{110}, 086403 (2013)}] to compute steady state properties of strongly correlated electronic systems out of equilibrium. The method can be considered as a non-equilibrium generalization of exact diagonalization based dynamical mean-field theory (DMFT). The key modification for the non-equilibrium situation consists in addressing the DMFT impurity  problem within an auxiliary system  consisting of the correlated impurity, $N_b$ uncorrelated bath sites and two Markovian environments (sink and reservoir). {Algorithmic improvements in the impurity solver allow to treat efficiently larger values of $N_b$ than previously in DMFT. This increases the accuracy of the results and is crucial for a correct description of the physical behavior of the system in the relevant parameter range including a semi-quantitative description of the Kondo regime.} To illustrate the approach we consider a monoatomic layer of correlated orbitals, described by the single-band Hubbard model, attached to two metallic leads. The non-equilibrium situation is driven  by a bias-voltage applied to the leads.  For this system, we investigate the spectral function and the steady state current-voltage characteristics in the weakly as well as in the strongly interacting limit.  {In particular we investigate the non-equilibrium behavior of quasi-particle excitations within the Mott gap of the correlated layer. We find for low bias voltage Kondo like behavior in the vicinity of the insulating phase. In particular we observe a splitting of the  Kondo resonance as a function of the bias voltage.}
\end{abstract}

\ifx\clength\undefined
\maketitle
\else
\nocomm
\fi

\section{Introduction}

The recent impressive experimental progress in tailoring different microscopically controlled quantum objects has prompted increasing interest in correlated systems out of  equilibrium. Of particular importance are correlated heterostructures\cite{an.ga.99, is.og.01,oh.mu.02, oh.hw.04, ga.ah.02, zh.wa.12}, quantum wires\cite{ta.de.97} and quantum dots\cite{go.sh.98, cr.oo.98} with atomic resolution, experiments in ultra cold atomic gases in optical lattices\cite{ra.sa.97, st.oe.12, ja.br.98, gr.ma.02, fa.sa.04}, as well as ultrafast laser spectroscopy\cite{iw.on.03, ca.de.04, pe.lo.06, fa.to.11}. 

The theoretical description and understanding of these experiments in particular and of complex strongly correlated systems in general presents major challenges  to theoretical solid state physics.
For this purpose different theoretical approaches have been developed. {\em For the equilibrium situation} one of the most powerful methods is dynamical mean-field theory (DMFT)\cite{ge.ko.96, Voll.10, me.vo.89}, which is a comprehensive, thermodynamically consistent and non-perturbative scheme. The only approximation in DMFT is the locality of the self-energy, which becomes exact in infinite dimensions, but usually it is a good approximation for two and three spatial dimensions. The key point of DMFT is to map the original problem onto a single impurity Anderson model (SIAM)\cite{ande.61} whose  parameters are determined self-consistently.  For this purpose, several classes of so-called impurity solvers were developed. Among them, the most powerful methods  are the numerical renormalization group (NRG) approach\cite{wils.75, kr.wi.80, bu.co.08}, Quantum Monte Carlo (QMC)\cite{hi.fy.86,ru.sa.05,we.co.06,gu.mi.11} and exact diagonalization (ED)\cite{ca.kr.94, si.ro.94}.

\begin{figure}[t]
 \includegraphics[width=\columnwidth]{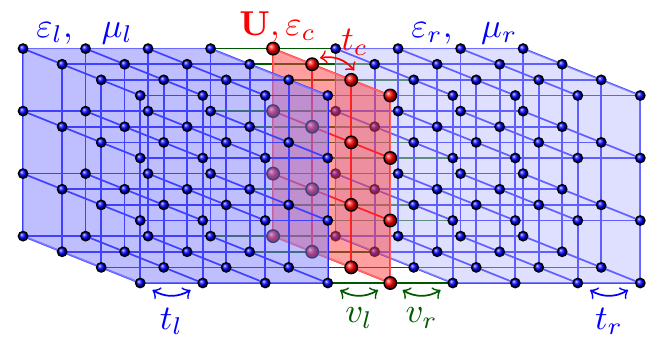}
\caption{(Color online) Schematic representation of the system, consisting of the correlated layer (red) with local Hubbard interaction $U$ and on-site energy $\varepsilon_c$, sandwiched between two semi-infinite metallic leads (blue), with on-site energies $\varepsilon_l$ and $\varepsilon_r$ respectively. The hopping between neighboring sites of the correlated layer is $t_c$, while the one for the left (right) lead is $t_l$ ($t_r$). Hybridization between the left (right) lead and the correlated layer is $v_l$ ($v_r$). A bias voltage $\Phi=\mu_l-\mu_r$ is applied between the leads.
}
\label{schematicp}
\end{figure}

Prompted by the success of DMFT for equilibrium systems, the approach was extended\cite{sc.mo.02u, fr.tu.06, free.08, jo.fr.08, ec.ko.09, okam.07, okam.08} to deal with time-dependent problems within the nonequilibrium Green's function approach originating from the works of Kubo\cite{kubo.57}, Schwinger\cite{schw.61}, Kadanoff, Baym\cite{ba.ka.61, kad.baym} and Keldysh\cite{keld.65}. Similar to the equilibrium case, also non-equilibrium DMFT is based on the solution of an appropriate (non-equilibrium) SIAM. Despite the fact that many approaches have been suggested to solve such impurity problems (see, e.g. Refs.~\onlinecite{sc.mo.02u,jo.fr.08,ec.ko.09,ec.ko.10,okam.08,me.an.06,ande.08,ro.pa.05,scho.09,wh.fe.04,da.ko.04,an.sc.05,kehr.05,ge.pr.07,ja.me.07,ha.he.07,di.we.10,he.fe.09,gr.ba.13,wo.mc.14}), not all of them are suited for non-equilibrium DMFT. In addition, many of these are only reliable for short times and cannot treat long time  behavior and accurately describe the steady state. Therefore, developing a non-perturbative impurity solver, which can treat reliably the steady state behavior of the SIAM is quite a challenge. 

{ The non-equilibrium approach, that will be presented in this paper, has its root in the exact diagonalization (ED)}-based DMFT (ED-DMFT). In equilibrium ED-DMFT one replaces the infinite bath by an auxiliary finite non-interacting electronic chain whose parameters are determined by a fit to the DMFT hybridization function $\Delta$. This cannot be trivially extended  to the steady state situation.
 First of all, due to the fact that the auxiliary system is finite, there is no dissipation and a proper steady state is never reached. An additional technical aspect is that the spectrum of the auxiliary system is discrete and therefore, the fit in real frequencies is problematic. But only in the equilibrium case one can circumvent this problem by introducing a fit in Matsubara space.~\cite{footnote_DAC}
A possible solution to these problems was suggested by us in Refs.~\onlinecite{ar.kn.13, do.nu.14} with an approach which enables  direct access to the steady state properties of the correlated impurity problem. The basic idea is that in addition to a finite number of bath sites coupled to the impurity, as in equilibrium ED, two Markovian environments are introduced, which act as  particle sink and reservoir. This auxiliary model represents an open quantum system with dissipative dynamics, which allows to properly describe steady state situations. The behavior of this auxiliary non-equilibrium impurity problem is described by a Lindblad master equation, which can be solved exactly by numerical approaches such as {full diagonalization \cite{ar.kn.13}}, non-Hermitian Krylov space~\cite{do.nu.14} or matrix product state (MPS) methods.~\cite{do.ga.15}
Its solution allows to determine both, the retarded and the Keldysh self-energies, which are required by the DMFT loop, with high accuracy. 
{Here, in particular, we apply the Krylov space approach of Ref. \onlinecite{do.nu.14} to solve the DMFT impurity problem. This yields a much better accuracy than in Ref.~\onlinecite{ar.kn.13}, which allows us to resolve the splitting of the quasi-particle resonance as a function of the bias voltage.}

The paper is organized as follows: In Sec. \ref{Model} we shortly introduce the Hamiltonian of the system, while in Secs. \ref{non-equilibrium_Green_functions} and \ref{DMFT} we give an overview over steady-state DMFT within the non-equilibrium Green's function  formalism. In Sec. \ref{imps} we discuss the auxiliary master equation approach, with focus on details of our implementation. Afterwards in Sec. \ref{Results} we present our results for a simple correlated interface. In particular, in Sec. \ref{Convergence} we benchmark the accuracy, while in Secs. \ref{Steady-state_current} and \ref{spectral_function} the steady state current and spectral functions are investigated, respectively. Finally in Sec. \ref{Conclusions} we give concluding remarks and an outlook.

\section{Model and Method}\label{Model_and_Method}

\subsection{Model}\label{Model}

To illustrate the approach we consider {a minimalistic model for transport across a correlated interface (see \Fig{schematicp})}, which consists of a correlated infinite and transitionally invariant layer (c), with local Hubbard interaction $U$, on-site energy $\varepsilon_c=-U/2$  and nearest-neighbor hopping amplitude $t_c$, sandwiched between two semi-infinite metallic leads ($\alpha=l,r$), with on-site energies $\varepsilon_\alpha$ and nearest-neighbor hopping amplitudes $t_\alpha$. The leads are semi-infinite and translationally invariant in the $xy$ plane (parallel to the correlated layer). The hybridization between lead $\alpha$ and the correlated layer is $v_\alpha$ (See Fig. \ref{schematicp}). A bias voltage $\Phi$ is applied between the leads. The Hamiltonian reads

\begin{equation}
\label{Hamiltonian}
{\cal H}={\cal H}_c + \sum_{\alpha=l,r}{\cal H}_\alpha +{\cal H}_{\rm coup} \, .
\end{equation}
Here 
\begin{equation}
\label{Hamiltonian_c}
{\cal H}_c= -t_c \sum_{\langle ij\rangle, \sigma}c_{i\sigma}^\dagger c_{j\sigma}^{\phantom\dagger} +U\sum_i n_{i\uparrow}n_{i_\downarrow} +\varepsilon_c\sum_{i,\sigma}n_{i\sigma} 
\end{equation}
describes the correlated layer. $ \langle i,j\rangle$ stands for neighboring $i$ and $j$ sites, $c_{i,\sigma}^\dagger$ creates an electron at the $i$-th site of the correlated layer with spin $\sigma=\uparrow,\downarrow$ and $n_{i\sigma}=c_{i \sigma}^\dagger c_{i \sigma}^{\phantom\dagger}$ denote the corresponding occupation-number operators. The leads are described by the Hamiltonian
\begin{equation}
\label{Hamiltonian_alpha}
{\cal H}_\alpha= -t_\alpha \sum_{\langle i j\rangle,  \sigma}c_{\alpha i \sigma }^\dagger c_{\alpha j \sigma }^{\phantom\dagger} 
+\varepsilon_\alpha\sum_{i \sigma}c_{\alpha i\sigma}^\dagger c_{\alpha i \sigma }^{\phantom\dagger} \, .
\end{equation}
Here  $c_{ \alpha i \sigma}^\dagger$ creates an electron at $i$-th site of the lead $\alpha$. An applied bias voltage $\Phi$ shifts the energies $\varepsilon_\alpha$ and chemical potentials $\mu_\alpha$ of the leads in opposite directions by the amount $\Phi/2$. Finally, 
\begin{equation}
{\cal H}_{\rm coup}={-}\sum_{\langle i j\rangle,  \alpha, \sigma} v_\alpha\left(c_{i\sigma}^\dagger c_{ \alpha j\sigma}^{\phantom\dagger} + h.c.\right)
\end{equation}
describes the hybridization between the correlated layer and leads. The hopping $v_\alpha$ takes place between neighboring sites of the lead and the correlated layer. 

{Previously similar models with many correlated layers were also investigated in Refs.~\onlinecite{okam.07, okam.08, ek.we.13, ma.am.15}~. In Refs.~\onlinecite{okam.07, okam.08} steady state behavior, while in Refs.~\onlinecite{ek.we.13, ma.am.15} full time evolution were investigated. For this purpose the authors used DMFT (Refs.~\onlinecite{okam.07, okam.08, ek.we.13}) and time-dependent Gutzwiller approximation (Ref.~\onlinecite{ma.am.15}). In the Refs.~\onlinecite{okam.07, okam.08} the impurity problem is treated by an  equation-of-motion approach with a suitable decoupling scheme for the higher order Green's functions, while in Ref.~\onlinecite{ek.we.13} the  non-crossing approximation is invoked. On the other hand, our treatment of the impurity solver is controlled and can achieve extremely accurate results \cite{do.ga.15} with a moderate number of bath sites.
}

\subsection{Steady-state non-equilibrium Green's functions}\label{non-equilibrium_Green_functions}

We consider an initial situation in which at  times $\tau<0$ the leads are disconnected from the correlated layer and all three parts  of the system ($l$, $c$, $r$) are in equilibrium with different values for the chemical potential $\mu_l=\varepsilon_l$, $\mu_c=\varepsilon_c$ and $\mu_r=\varepsilon_r$, respectively.

Due to the fact that the system is transitionally invariant in the $xy$ plane, it is more convenient to perform a Fourier transformation and express the Green's functions in terms of the momentum ${\bf k}_{||}=(k_x,k_y)$. The  retarded equilibrium Green's function for the disconnected non-interacting central layer reads
\begin{equation}
g_0^R (\omega, {\bf k}_{||})=\frac{1}{\omega+ i 0^{+} -\varepsilon_c -E_c( {\bf k}_{||})} \, ,
\label{g0R}
\end{equation}
with $E_c( {\bf k}_{||})=-2t_c (\cos k_x  + \cos k_y)$. On the other hand, the Green's functions for the edge layers of the left ($\alpha=l$) and the right ($\alpha=r$) lead, when they are disconnected from the central layer can be  expressed as\cite{hayd.80, po.no.99.sm, po.no.99.ms}
\begin{eqnarray}
&&g_\alpha^R(\omega, {\bf k}_{||})= \frac{\omega-\varepsilon_\alpha -E_\alpha({\bf k}_{||})}{2t_\alpha^2}  \nonumber \\
&&\hspace{1.625cm}- i\frac{\sqrt{4t_\alpha^2-(\omega-\varepsilon_\alpha-E_\alpha({\bf k}_{||}))^2}}{2t_\alpha^2} \, ,
\label{galphaR}
\end{eqnarray}
with $E_\alpha( {\bf k}_{||})=-2t_\alpha (\cos k_x  + \cos k_y)$. The sign of the square-root for negative argument must be chosen such that the Green's function has the correct $1/\omega$ behavior for $|\omega|\to \infty$. To investigate the system out of the equilibrium, we need to work within the Keldysh Green's function formalism.~\cite{kad.baym, schw.61, keld.65, ha.ja, ra.sm.86}  Therefore, as a starting point, we need the corresponding non-interacting, disconnected Keldysh components. Since the disconnected systems are separately in equilibrium, we can obtain these from the retarded ones via the fluctuation dissipation theorem \cite{ha.ja}
\begin{equation}
g_\alpha^K(\omega, {\bf k}_{||})= 2i(1-2f_\alpha(\omega))\;\Imm g_{\alpha}^R(\omega,{\bf k}_{||}) \, .
\label{galphaK}
\end{equation}
Here, $f_\alpha(\omega)$ is the Fermi distribution for chemical potential $\mu_\alpha$ and temperature $T_\alpha$. For the non-interacting isolated central layer, the inverse Keldysh Green's function $[\underline g_0^{-1}(\omega, {\bf k}_{||})]^K$ is infinitesimal  and can be neglected in a steady state in which  the layer is connected to the leads. In our notation, we use an underline to denote   block matrices within the non-equilibrium Green's function (Keldysh) formalism:
\begin{equation}
\underline X =\left(
\begin{array}{cc}
X^R & X^K \\
0   & X^A 
\end{array}
\right) 
\label{Keldysh-str}
\end{equation}
with $X^A=(X^R)^\dagger$.
At time $\tau=0$ the leads get connected to the correlated layer. After a sufficiently long time a steady state is reached. The latter is expected to exist and to be unique unless the system has bound states. Our goal is to investigate its properties under the bias voltage $\Phi$. 

Since the steady state is time-translation invariant, we can Fourier transform in time and express all Green's functions in terms of a real frequency $\omega$. The Green's function for the correlated  layer, when connected with the leads, can be expressed via Dyson's equation
\begin{equation}
{\underline G}^{-1}(\omega, {\bf k}_{||}) ={\underline G}_0^{-1}(\omega, {\bf k}_{||}) - {\underline{\Sigma}}(\omega, {\bf k}_{||}) \, ,
\label{Dyson}
\end{equation}
where ${\underline{\Sigma}}(\omega, {\bf k}_{||})$ is the is self-energy of the correlated layer. The Green's function of the non-interacting non-equilibrium system ${\underline G}_0 (\omega, {\bf k}_{||})$ in turn can be expressed  as
\begin{align}
{\underline G}_0^{-1}(\omega, {\bf k}_{||}) &= {\underline g}^{-1}_0(\omega, {\bf k}_{||}) - \sum_{\alpha=l,r}v_\alpha^2~{\underline g}_\alpha(\omega, {\bf k}_{||})
\,,
\label{G0}
\end{align}
where ${\underline g}_0(\omega, {\bf k}_{||})$ is the Green's function of the non-interacting decoupled layer, i.e. $U=0,v_{\alpha}=0$ and the components of the Green's function of the isolated leads ${\underline g}_\alpha(\omega, {\bf k}_{||})$ are given in \eq{galphaR} and \eq{galphaK}. Note that all quantities are underscored, i.e. they are Keldysh block matrices.

\begin{figure}[t]
\includegraphics[width=\columnwidth]{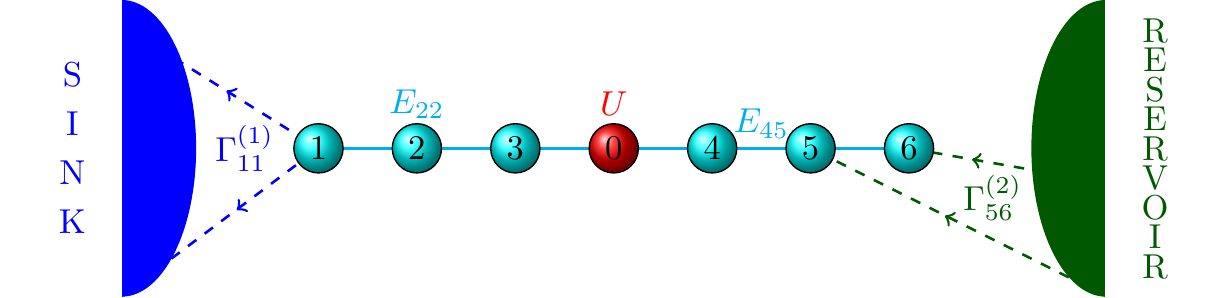}
\caption{(Color online) Sketch of the auxiliary open quantum impurity problem consisting of the impurity site at position $i=0$ (red circle), $N_b=6$ bath sites (cyan circles) and two Markovian environments sink (blue) and reservoir (green). Parameters $E_{ij}$ and 
$\Gamma_{ij}$ are explained in the main text. 
}
\label{Auxiliary}
\end{figure}

\subsection{Dynamical mean-field theory}\label{DMFT}

As usual, to obtain the self-energy ${\underline{\Sigma}}(\omega, {\bf k}_{||})$  is the difficult step in the calculation of ${\underline G}(\omega, {\bf k}_{||})$ and of various steady state properties of the system. As there is no closed expression for it, one has to resort to some approximation. Here,  we employ DMFT~\cite{ge.ko.96, Voll.10, me.vo.89,sc.mo.02u,fr.tu.06,okam.07,ar.kn.13} in its nonequilibrium, time-independent version. In this approach the self-energy is approximated by a local quantity ${\underline{\Sigma}}(\omega, {\bf k}_{||}) = {\underline{\Sigma}}(\omega)$ which can be determined by solving a (non-equilibrium) quantum impurity model with the same Hubbard interaction $U$ and on-site energy $\varepsilon_c$ coupled to a self-consistently determined bath. The latter is specified by its hybridization function obtained as
\begin{equation}
\underline{\Delta}(\omega)=\underline{g}_0^{-1}(\omega) - \underline{G}^{-1}_{\rm loc}(\omega)-\underline{\Sigma}(\omega)\, , 
\label{Delta}
\end{equation}
where $\underline{g}_0^{-1}(\omega)$ is the non-interacting Green's function of the disconnected impurity (i.e. of a single correlated site) and 
\begin{equation}
\underline{G}_{\rm loc}(\omega)=\int\limits_{\rm BZ} \frac{d{\bf k}_{||}}{(2\pi)^2}\underline{G}(\omega,{\bf k}_{||}) \, .
\label{G_loc}
\end{equation}

The self-consistent DMFT loop works similarly to the equilibrium case, except that in the present case the Green's functions are $2\times 2$ block matrices\cite{fr.tu.06, okam.07, okam.08}: One  starts with an initial guess for the self-energy $\underline{\Sigma}(\omega)$, then based on Eqs. \eqref{g0R}-\eqref{G_loc} calculates the bath hybridization function $\underline{\Delta}(\omega)$. We then evaluate the corresponding auxiliary Green's functions $\underline{G}_{{\rm aux},0}(\omega)$ and $\underline{G}_{\rm aux}(\omega)$, in the non-interacting and in the interacting case, respectively. The solution of the impurity problem is, as usual, the bottleneck of DMFT. Our scheme consists, as outlined in detail in Sec.~\ref{imps}, in replacing the impurity problem with an auxiliary one, which is as close as possible to the one described by \eqref{Delta} but is exactly solvable by numerical methods. The self-consistent loop is then closed by determining the new value of the self-energy 
\begin{equation}
\underline{\Sigma}(\omega)=\underline{G}_{{\rm aux},0}^{-1}(\omega)-\underline{G}_{\rm aux}^{-1}(\omega) \,.
\label{Self-Energy}
\end{equation}
We repeat this procedure until convergence is reached, i.e. until $\underline{G}_{\rm aux}(\omega) \approx \underline{G}_{\rm loc}(\omega)$. \cite{footnote_Convergence}

\subsection{Impurity solver: auxiliary master equation approach }\label{imps}

As already mentioned, the main obstacle of DMFT is the solution of the impurity problem. One widespread approach to approximate its solution for the equilibrium case is ED-DMFT, whereby one replaces the infinite bath with an auxiliary finite one. However, this approach cannot be used straightforwardly in a nonequilibrium steady-state case, as this cannot be described correctly with a finite number of sites\cite{do.nu.14}. One way to overcome this problem, as some of us already suggested in Refs.~\onlinecite{ar.kn.13, do.nu.14}, is to introduce, in addition to the finite number $N_b$ of bath sites which are coupled to the impurity in the form of two chain segments, two Markovian environments, which can be seen as a particle sink and reservoir, respectively (see Fig. \ref{Auxiliary}). This makes the impurity model effectively infinitely large, which is necessary in order to be able to reach a  steady state. Our strategy, similar to the equilibrium ED-DMFT case, is to choose the parameters of the auxiliary model so as to
provide an optimal fit to the  bath hybridization function \eqref{Delta}. 

The dynamics of this auxiliary impurity model, is described by the Lindblad quantum master equation, which controls the time ($\tau$) dependence of the reduced density matrix $ \rho$ of the model\cite{br.pe,carmichael1}
\begin{equation}
\frac{d}{d\tau} \rho={{ {\cal L}}} \rho  \, , 
\label{Lindblad_Eq}
\end{equation}
where 
\begin{equation}
{{{\cal L}}}= {{{\cal L}}}_H+{{{\cal L}}}_D
\label{Linblad_Superoperator}
\end{equation}
is a Lindblad superoperator, which consist of two terms: a unitary contribution $ {{{\cal L}}}_H$ and a dissipative one ${{{\cal L}}}_D$. 

The unitary contribution
\begin{equation}
{{{\cal L}}}_H \rho = - i [{{\cal H}}_{\rm aux}, \rho] \, ,
\label{L_H}
\end{equation}
is generated by the auxiliary Hamiltonian
\begin{equation}
{{\cal H}}_{\rm aux}=\sum_{i,j=0{, \sigma}}^{N_b}E_{ij} d_{i\sigma}^\dagger d_{j\sigma}^{\phantom\dagger} + U\, n_{0\uparrow}^d n_{0\downarrow}^d\, ,
\label{Haux}
\end{equation}
where $d_{i\sigma}^\dagger$ creates a particle with spin $\sigma$ at the impurity ($i=0$) or at a bath site $(i=1, \ldots, N_b)$. $n_{0\sigma}^d=d_{0\sigma}^\dagger d_{0\sigma}^{\phantom\dagger}$ is the occupation number operator for particles at the impurity-site with spin $\sigma$. $E_{00}=\varepsilon_c$, while all other $E_{ij}$ are parameters used to fit $\underline \Delta(\omega)$, whereby one can restrict to  on-site and nearest neighbor (n.n.) terms only (see Fig. \ref{Auxiliary}). The non-unitary  (dissipative) term  
\begin{eqnarray}
&&{{{\cal L}}}_D \rho =2 \sum_{i,j=0}^{N_b}\sum_\sigma \Bigg[\Gamma_{ij}^{(1)}\left(d_{i\sigma}^\nag\rho\  d_{j\sigma}^{\dag}
-\frac{1}{2}\{\rho,d_{j\sigma}^{\dag}d_{i\sigma}^\nag\}\right) \nonumber \\ 
&&\hspace{0.8cm}+\Gamma_{ij}^{(2)}\left(d_{j\sigma}^{\dag}\rho\  d_{i\sigma}^\nag-\frac{1}{2}\{\rho,d_{i\sigma}^\nag d_{j\sigma}^{\dag}\}\right)\Bigg]\,
\label{H_D} 
\end{eqnarray}
describes  the coupling to a Markovian environment. The dissipation matrices ${\bf\Gamma}^{(\kappa)}$, $\kappa=1,2$ (with matrix elements $\Gamma^{(\kappa)}_{ij}$) are Hermitian and positive semidefinite \cite{br.pe} and are again used as fit parameters. In order to fix the large-$\omega$ behavior of $\underline \Delta$, all $\Gamma_{ij}^{({\kappa})}$ with at least one index on the impurity must vanish. On the other hand, in contrast to $\bf E$, ${\bf\Gamma}^{(\kappa)}$ are not restricted to n.n. terms. This is of great advantage for the fit, as discussed below in \se\ref{Convergence}. 

To carry out the self-consistent DMFT loop we need to evaluate both the non-interacting and the interacting Green's functions of the auxiliary model. First, the non-interacting calculation ($U=0$), which is fast in comparison to the interacting one, produces the bath hybridization function $\underline{\Delta}_{\rm aux}(\omega)$ of the auxiliary impurity model, which is fitted to \eqref{Delta} in order to obtain the optimal parameters ${\bf E}$ and ${\bf \Gamma}^{(\kappa)}$. These are used in the interacting model in order to determine  the self-energy $\underline{\Sigma}(\omega)$, which is then inserted in \eqref{Dyson}.

A convenient way to solve the auxiliary problem is to rewrite Eq. \eqref{Lindblad_Eq}, expressed by superoperators, into a standard operator problem\cite{pros.08, dz.ko.11, schm.78, ha.mu.08}. For this purpose one enlarges the original  Fock space, spanned by the operators ($d_{i,\sigma}^{\phantom\dagger}/d_{i,\sigma}^\dagger$), by doubling the number of levels via so-called tilde operators  $\tilde d_{i,\sigma}^{\phantom\dagger}/\tilde d_{i,\sigma}^\dagger$.  In addition one introduces a so-called left vacuum
\begin{equation}
| I \rangle =\sum_{S}(-1)^{N_S} |S \rangle \otimes | \tilde S \rangle \, ,
\label{Levt_Vacuum}
\end{equation}
where $| S \rangle$ are many body states of the original Fock space, $| \tilde S \rangle$ the corresponding ones of the tilde space\cite{dz.ko.11} and $N_S$ the number of particles in $S$. In this formalism the reduced density operator is mapped onto the state  vector
\begin{equation}
|\rho(\tau) \rangle = \rho| I \rangle \, ,
\label{rho}
\end{equation}
and the Lindblad equation is mapped onto a Schr\"odinger-type  equation\cite{dz.ko.11}
\begin{equation}
\frac{d}{d\tau} |\rho(\tau) \rangle =L|\rho(\tau) \rangle \, ,
\label{Lindblad_Eq_new}
\end{equation}
where 
\begin{equation}
L=L_0 + L_I
\label{Lindblad_L}
\end{equation}
is an ordinary operator in the augmented space. Its non-interacting part $L_0$ reads 
\begin{equation}
 iL_0=\sum_\sigma \left( {\bf d}_\sigma^\dagger {\bf h} {\bf d}_\sigma^{\phantom\dagger} - {\rm Tr}({\bf E} + i{\bf \Lambda})\right)
\label{L_0}
 \end{equation}
where $\rm Tr$ denotes the matrix trace, and
\begin{equation}
{\bf d}_\sigma^\dagger=\left(d_{0,\sigma}^\dagger, \ldots , d_{N_b,\sigma}^\dagger,  \tilde d_{0,\sigma}^{\phantom\dagger}, \ldots , \tilde d_{N_b,\sigma}^{\phantom\dagger}\right) \, . 
\end{equation}
is a vector of creation/annihilation  operators and the matrix $\bf h$ is given by 
\begin{equation}
{\bf h}=\left(
\begin{array}{cc}
{\bf E} +i {\bf \Omega} & 2{\bf \Gamma}^{(2)} \\
- 2{\bf \Gamma}^{(1)} & {\bf E} -i {\bf \Omega}
\end{array}
\right)
\label{Matrix_h}
\end{equation}
with 
\begin{equation}
{\bf \Lambda}={\bf \Gamma}^{(2)} + {\bf \Gamma}^{(1)} \, , \quad {\bf \Omega}={\bf \Gamma}^{(2)} - {\bf \Gamma}^{(1)} \, .
\end{equation}
Its interacting part has the form
\begin{equation}
i L_I = U n_{0\uparrow} n_{0 \downarrow}
- U \tilde n_{0\uparrow}\tilde n_{0\downarrow}\, ,
\end{equation}
with $\tilde n_{0\sigma} :=\tilde d^{\dagger}_{0\sigma} \tilde d^{\phantom\dagger}_{0\sigma}$. To evaluate Green's functions, one needs to calculate expectation values of the form
\begin{equation}
G_{BA}=-i {\rm tr}_{\cal U}  \big(B(\tau_2)  A(\tau_1)  \rho_{\cal U}(\tau_1)\big) \, ,
\label{Gree_BA}
\end{equation}
where $ \rho_{\cal U}(\tau_1)$ is the density operator of the ``universe'' $\cal U$ composed of the ``system'' (the chain in Fig.~\ref{Auxiliary}) and the Markovian environment and $\tr_{\cal U}=\tr \otimes \tr_E$ is the trace over the ``universe'', which is the tensor product of the trace over the ``system'' ($\tr$) and the trace over the environment ($\tr_E$).  After straightforward calculations we obtain (more details see Ref. \onlinecite{do.nu.14}) for the non-interacting retarded Green's function 
\begin{equation}
{\bf G}_{{\rm aux},0}^R=\left(\omega -{\bf E} + i {\bf \Lambda} \right)^{-1}
\label{G0_R}
\end{equation}
while its Keldysh part reads
\begin{equation}
{\bf G}_{{\rm aux},0}^K=2 i{\bf G}_{{\rm aux},0}^R {\bf \Omega}{\bf G}_{{\rm aux},0}^A \, .
\label{G0_K}
\end{equation}
Therefore, we obtain the following expressions for the retarded auxiliary hybridization function 
\begin{equation}
\Delta_{\rm aux}^R=\omega - \varepsilon_c - \frac{1}{[{\bf G}_{{\rm aux},0}^R]_{00}}
\label{Delta_R}
\end{equation}
and its Keldysh part
\begin{equation}
\Delta_{\rm aux}^K=\frac{[{\bf G}_{{\rm aux},0}^K]_{00}}{|[{\bf G}_{{\rm aux},0}^R]_{00}|^2}
\label{Delta_K}\;.
\end{equation}
Here $X_{00}$ denotes the $00$ element (i.e. the one on the impurity) of the matrix $\bf X$. 

To calculate the  impurity Green's function for the interacting system we use Krylov-space based exact diagonalization. A full diagonalization is prohibitive for $N_B\gtrsim 3$ due to the fact that the Hilbert space is exponentially large. Particle conservation translates here into conservation of ${N_\sigma-\tilde N_\sigma}$. To calculate the steady state $|\rho_\infty \rangle $ we use an Arnoldi time evolution\cite{kn.ar.11.ec},  while for the calculation of Green's functions we employ the two-sided Lanczos algorithm.\cite{do.nu.14} 

To obtain the retarded and the Keldysh Green's functions we use the following relations (for details see Ref.~\onlinecite{do.nu.14}): 
\begin{eqnarray}
&&{\bf G}^R= {\bf G}^{>+} +{{\bf G}^{<+}}^\dagger\\
\label{GR}
&&{\bf G}^K= {\bf G}^{>+} +{\bf G}^{<+} - H.c. \, .
\label{GK}
\end{eqnarray}
The expressions for the greater and lesser  Green's functions are~\cite{footnote_Superscript} 
\begin{equation}
G_{{\rm aux},ij\sigma}^{>+}(\omega)=\sum_n\frac{\langle I| d_{i\sigma}^{\phantom\dagger} | R_{n}^{(+1)} \rangle \langle L_n^{(+1)}|d_{j\sigma}^{\dagger}|\rho_\infty \rangle}{\omega - l_n^{(+1)}}  \,,
\label{Ggreater}
\end{equation}
and 
\begin{equation}
G_{{\rm aux},ij\sigma}^{<+}(\omega)=\sum_n\frac{\langle I| d_{i\sigma}^\dagger | R_{n}^{(-1)} \rangle \langle L_n^{(-1)}|d_{j\sigma}^{\phantom\dagger}|\rho_\infty \rangle}{\omega - l_n^{(-1)}}  \,,
\label{Glesser}
\end{equation}
with the right ($|R_{n}^{(\pm 1)} \rangle$) and left ($\langle L_n^{(\pm 1)}|$) eigenstates and eigenvalues $l_n^{(\pm 1)}$ of the operator $L$ (Eq. \eqref{Lindblad_L}), in the sectors $N_\sigma-\tilde N_\sigma=\pm1$.

Once self-consistency in the DMFT loop (cf. Sec. \ref{DMFT}) is achieved, one can calculate desired physical quantities, e.g. the steady-state current. For this purpose we use the Meir-Wingreen expression\cite{ha.ja, me.wi.92, jauh} in its symmetrized  form, where summation over spin is implicitly assumed.
\begin{eqnarray}
&&\hspace{-0.5cm}J=i \int\limits_{\rm BZ} \frac{d{\bf k}_{||}}{(2\pi)^2}\int_{-\infty}^\infty \frac{d\omega}{2\pi}\left[\left(\gamma_l(\omega,{\bf k}_{||}) 
- \gamma_r(\omega,{\bf k}_{||})\right)G^<(\omega,{\bf k}_{||}) \right. \nonumber \\
&&\hspace{-0.25cm}\left.+\left(\bar\gamma_l(\omega,{\bf k}_{||}) - \bar\gamma_r(\omega,{\bf k}_{||})\right)\hspace{-0.1cm}\left(G^R(\omega,{\bf k}_{||}) - G^A(\omega,{\bf k}_{||})\right) \right] \, ,
\label{Current}
\end{eqnarray}
here $\gamma_\alpha(\omega,{\bf k}_{||})=-2v_\alpha^2 \Imm g_\alpha(\omega,{\bf k}_{||}) $ and $\bar\gamma_\alpha(\omega,{\bf k}_{||})=f(\omega -\mu_\alpha)\gamma_\alpha(\omega,{\bf k}_{||}) $.

\begin{figure}
\includegraphics[width=0.75\columnwidth]{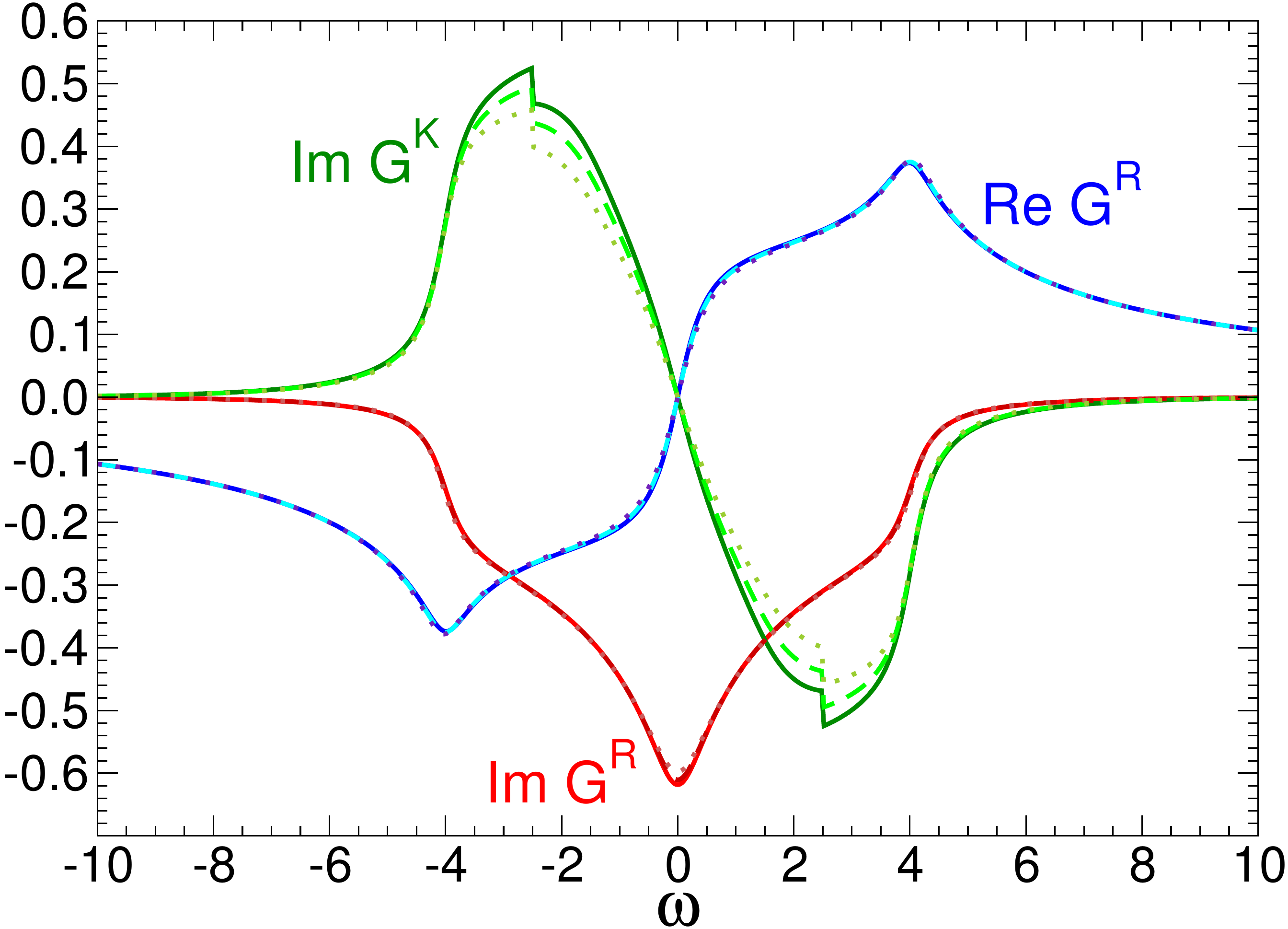} \\
\vspace{0.25cm}
\includegraphics[width=0.75\columnwidth]{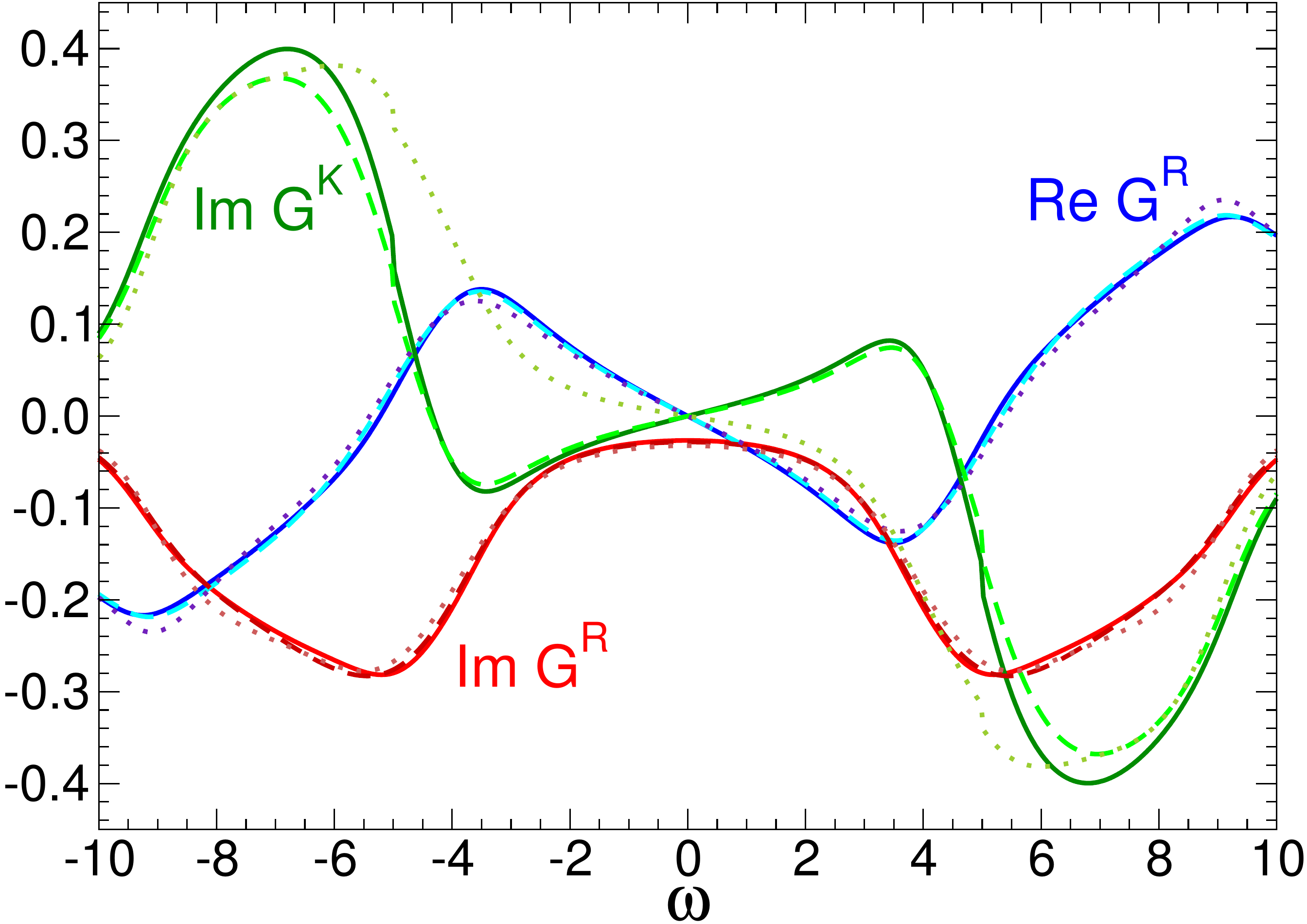}
\caption{(Color online) Retarded $G^R$ and Keldysh $G^K$ Green's functions for the correlated layer, for Hubbard interaction $U=2$ and bias voltage $\Phi=5$ (upper panel) and for $U=12$ and  $\Phi=10$ (lower panel). Other parameters are: $t_c=1$, $t_l=t_r=2.5$, $v_l=v_r=0.5$ ($2v_l$ is our unit of energy). Solid, dashed and dotted lines are obtained with $N_b=6$, $N_b=4$ and $N_b=2$ correspondingly.}
\label{Green_comp}
\end{figure}

\section{Results}\label{Results}

Here we present  results for the steady state properties of the system displayed in Fig.~\ref{schematicp}  consisting of a correlated layer, with Hubbard interaction $U$ and on-site energy $\varepsilon_c=-U/2$, coupled to two metallic leads. We restrict to the particle-hole symmetric case. The hopping inside the correlated layer $t_c$ is taken equal to $1$ unless stated otherwise, while the hopping inside the leads is $t_l=t_r=2.5$. The hybridizations between leads and the correlated layer are  $v_l=v_r=0.5$ and $2v_l$ is used as unit of energy. The applied bias voltage $\Phi$ enters the values of the onsite energies and the chemical potentials as $\varepsilon_{l/r}=\mu_{l/r}=\pm \Phi/2$. All results presented below are calculated for zero temperature in the leads ($T_l=T_r=0$), see \eq{galphaK}. Similar models have been studied, e.g.  in Refs.~\onlinecite{okam.07,okam.08,kn.li.11,ma.am.15,am.we.12}.

\subsection{Convergence with respect to the number of auxiliary bath sites $N_b$}\label{Convergence}

\begin{figure}
\includegraphics[width=0.75\columnwidth]{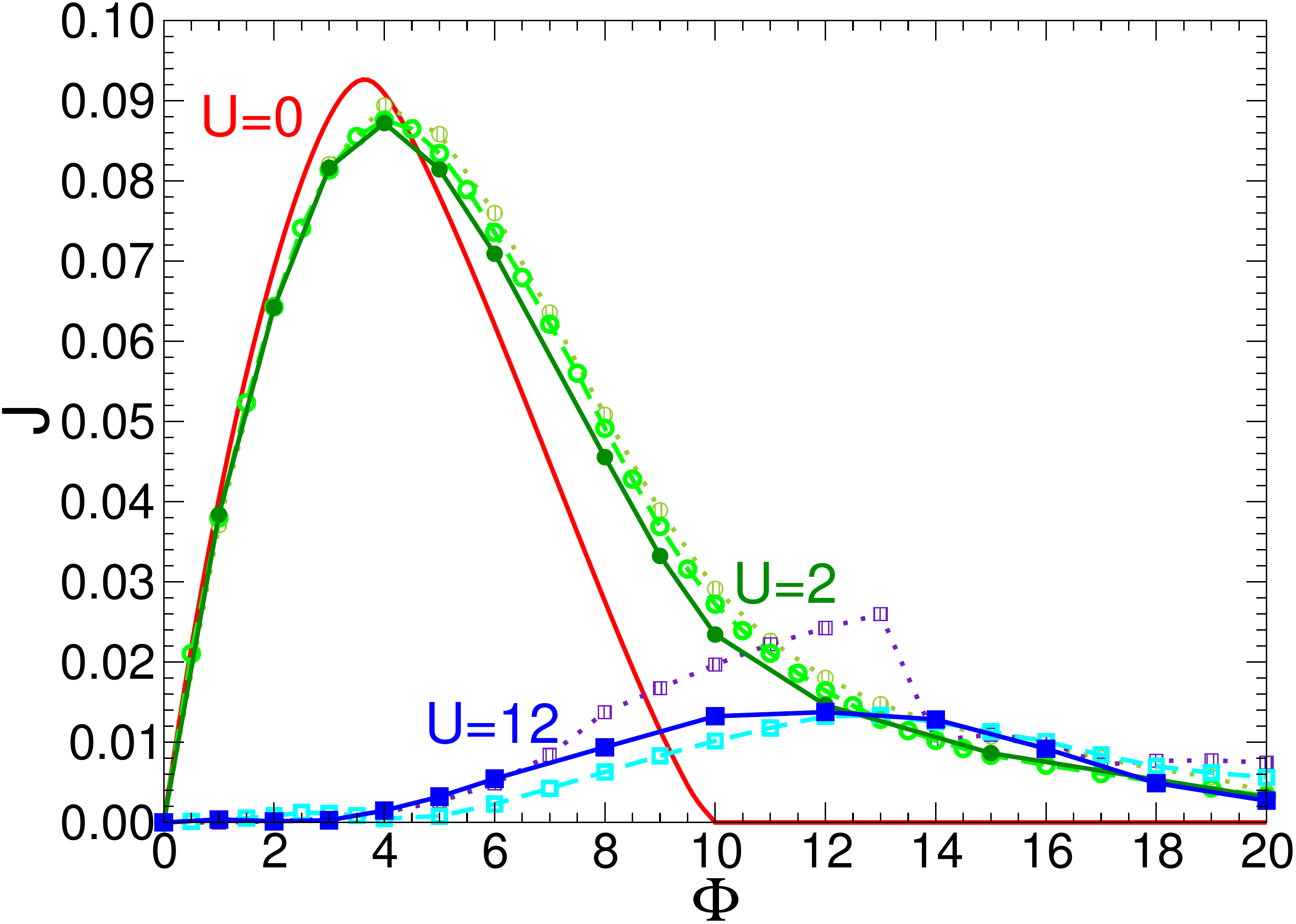}
\caption{(Color online) Current $J$ vs bias voltage $\Phi$ for different $U$. The solid red curve represents the $U=0$ result. The three curves peaked at about $\Phi=4$ are for $U=2$ and the remaining three curves show the $U=12$ results. Solid, dashed and dotted lines  correspond to $N_b=6$, $N_b=4$ and $N_b=2$, respectively. Other parameters are as in Fig.~\ref{Green_comp}.}
\label{Current_vs_Phi}
\end{figure}

First we investigate how the number of bath sites $N_b$ of the auxiliary impurity problem influences the results. We compare calculations for the Green's functions (Fig.~\ref{Green_comp}) and for the current (Fig.~\ref{Current_vs_Phi}), obtained with  $N_b=2,4,6$. We find that the retarded component is well converged already for $N_{b}=4$ even for $U=12$. For the Keldysh Green's functions the convergence in terms of $N_b$ is reasonable, but not as fast as for the retarded Green's function. Correspondingly, it is not surprising that also the current voltage characteristics exhibit a fairly good convergence (See Fig.~\ref{Current_vs_Phi}). On the whole, the convergence for weak interaction ($U=2$) is faster than for strong interaction ($U=12$). 

These results indicate that  $N_b=4$ bath sites already produce reasonable results away from the Kondo regime. Therefore, in view of the exponential increase of the numerical effort with $N_{b}$, we mainly restrict the following discussion to $N_{b}=4$. Only to discuss the low energy Kondo physics we will present results with  $N_{b}=6$. The reason for the rapid convergence in $N_{b}$ is due to the fact that the number of Lindblad parameters increases quadratically with $N_{b}$, in contrast to the energy and hopping parameters $\bf E$. It is, thus, important, to consider also long-ranged ${\bf \Gamma}$ terms (cf. Ref.~\onlinecite{do.nu.14})

\begin{figure}
 \includegraphics[width=0.75\columnwidth]{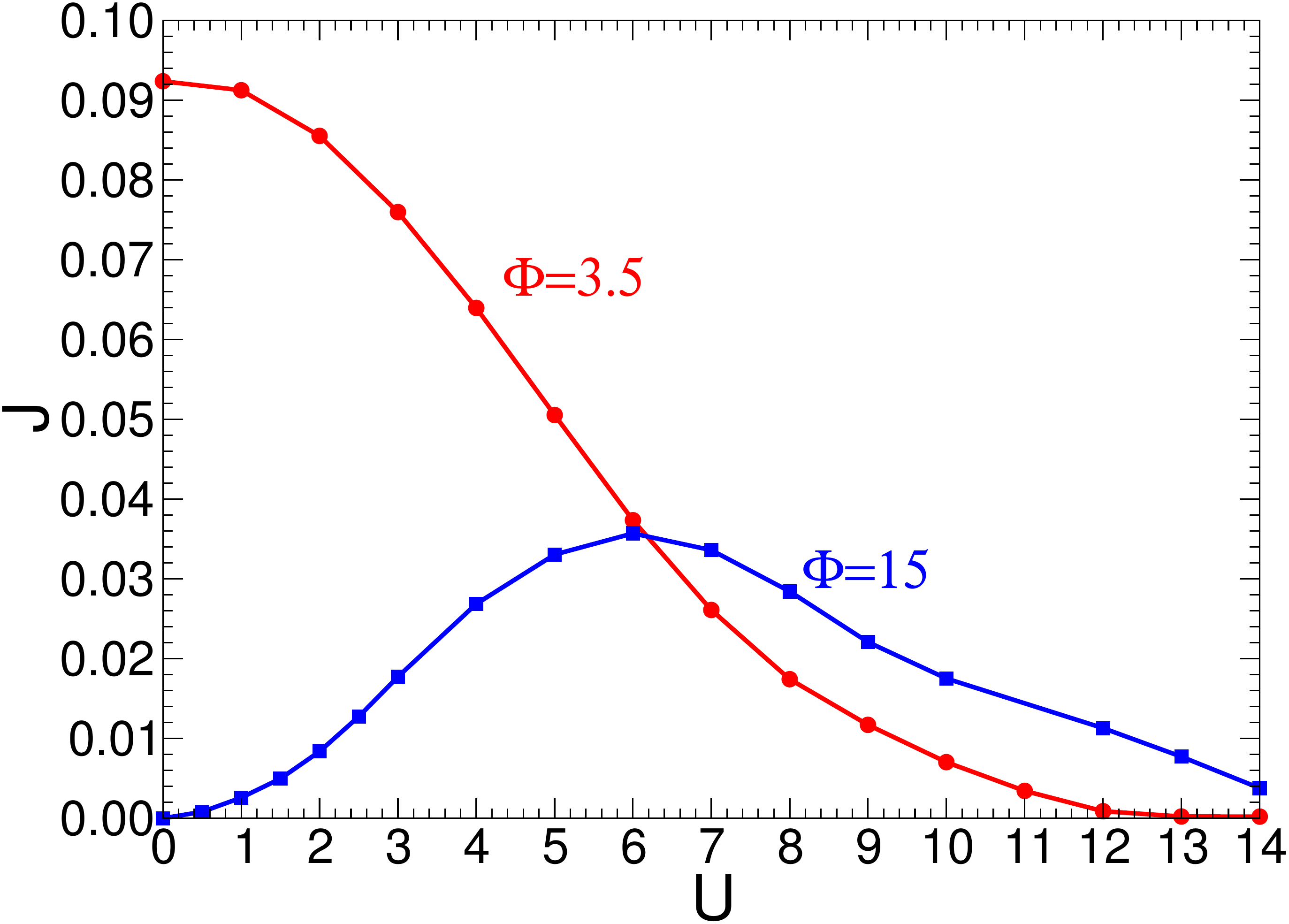}
\caption{(Color online) Current $J$ vs interaction $U$ for two values of $\Phi$. Calculations are performed for $N_b=4$. Other parameters are as in Fig.~\ref{Green_comp}. }
\label{Current_vs_U}
\end{figure}

\begin{center}
\begin{figure*}
\subfigure[]{
\label{Aomega_U2_color1}
\begin{minipage}[b]{0.35\textwidth}
\centering \includegraphics[width=1\textwidth]{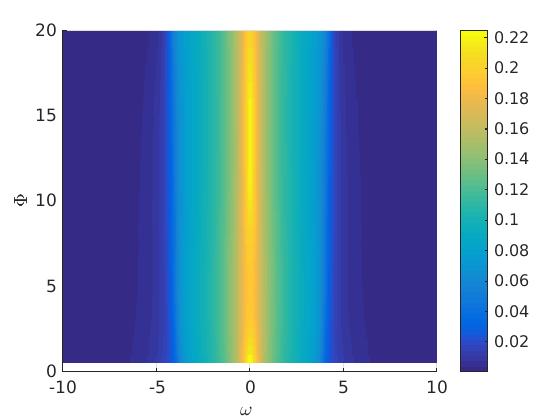}
\end{minipage}}
\hspace{0.05\textwidth}
\subfigure[]{
\label{Aomega_U12_color1}
\begin{minipage}[b]{0.35\textwidth}
\centering \includegraphics[width=1\textwidth]{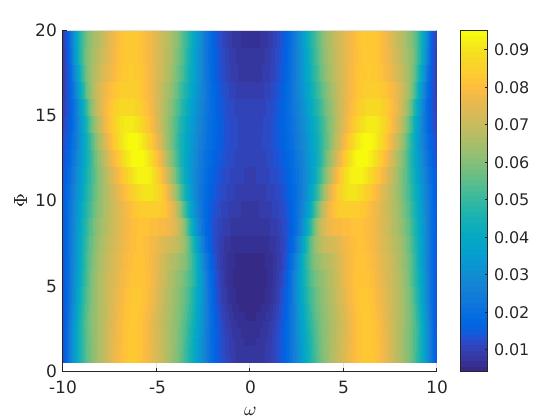}
\end{minipage}}
\\
\caption{(Color online) Single particle spectral function for $N_b = 4$, different values of $\Phi$, $U = 2$ (a) and $U = 12$ (b). Other parameters are as in Fig.~\ref{Green_comp}.}
\label{Aomega_fix_U}
\end{figure*}
\end{center}

\subsection{Steady-state current}\label{Steady-state_current}

In this section we discuss the steady-state current in detail. The results for the current as a function of bias voltage are presented in Fig~\ref{Current_vs_Phi} for different values of the Hubbard interaction $U$. In the non-interacting case ($U=0$) particles pass the interface without scattering and therefore the momentum ${\bf k}_{||}$ is conserved. Correspondingly, the problem becomes one-dimensional and the current  vanishes for bias voltages larger than the one-dimensional bandwidth, i.e. for $W_{z}=4t_{l/r} = 10$, which is corroborated by  Fig.~\ref{Current_vs_Phi}. For nonzero interaction different ${\bf k}_{||}$ are mixed due to scattering and thus all states of the leads are possible final states. Subsequently, the current vanishes for bias voltages larger than the three-dimensional bandwidth, i.e. $W=3\cdot W_{z}=30$. In equilibrium, an isolated two-dimensional Hubbard layer is in the metallic phase for weak interaction. As can be seen from Fig. \ref{Current_vs_Phi}, in {this case }($U=2$) the current displays, as expected, a metallic behavior, i.e. a linear increase of the current for small voltages. The overall shape is similar to the $U=0$ case, however, with a longer tail at large $\Phi$ due to the scattering mechanism discussed above. For strong interaction ($U=12$) {an isolated two-dimensional Hubbard layer is a Mott insulator, but in our model there is no insulating phase due to the hybridization to the non-interacting leads. Therefore, strictly speaking  the current is always linear in $\Phi$ for $\Phi\to 0$. Nevertheless, due to the vicinity of the Mott insulator the current is strongly suppressed.~\cite{footnote_Tiny_Current} A similar behavior also was observed in Refs.~\onlinecite{okam.07, okam.08}~.} On the other hand, for higher bias voltages ($\Phi \gtrsim 12$) the picture is reversed and the current is more suppressed for $U=2$ than for $U=12$.

We investigate this issue in detail  and plot the current as a function of Hubbard interaction $U$ for low (${\Phi=3.5}$) and high (${\Phi=15}$) bias voltage in Fig.~\ref{Current_vs_U}. For the low voltage case, we find a monotonic {Gaussian} decrease. The origin of the current reduction with increasing $U$ are back-scattering processes that reduce  the transmission coefficient. For high bias voltage, in the region where the current is zero for $U=0$, the current first increases with increasing interaction,  reaches its maximum at approximately $U\simeq 6$ and then decreases again. \cite{footnote_Max_Shift} Qualitatively this can be explained by the fact that  {there are two competing effects as a function of $U$. On the one hand, with increasing $U$ the transport increases due to scattering to different ${\bf k}_{||}$ as discussed above, which enhances the current, but on the other hand, large $U$ means increased backscattering which suppresses transport across the correlated layer. For high bias voltages and weak interactions the first effect dominates due to the finite bandwidth of the leads.}

\begin{figure}
\includegraphics[width=0.75\columnwidth]{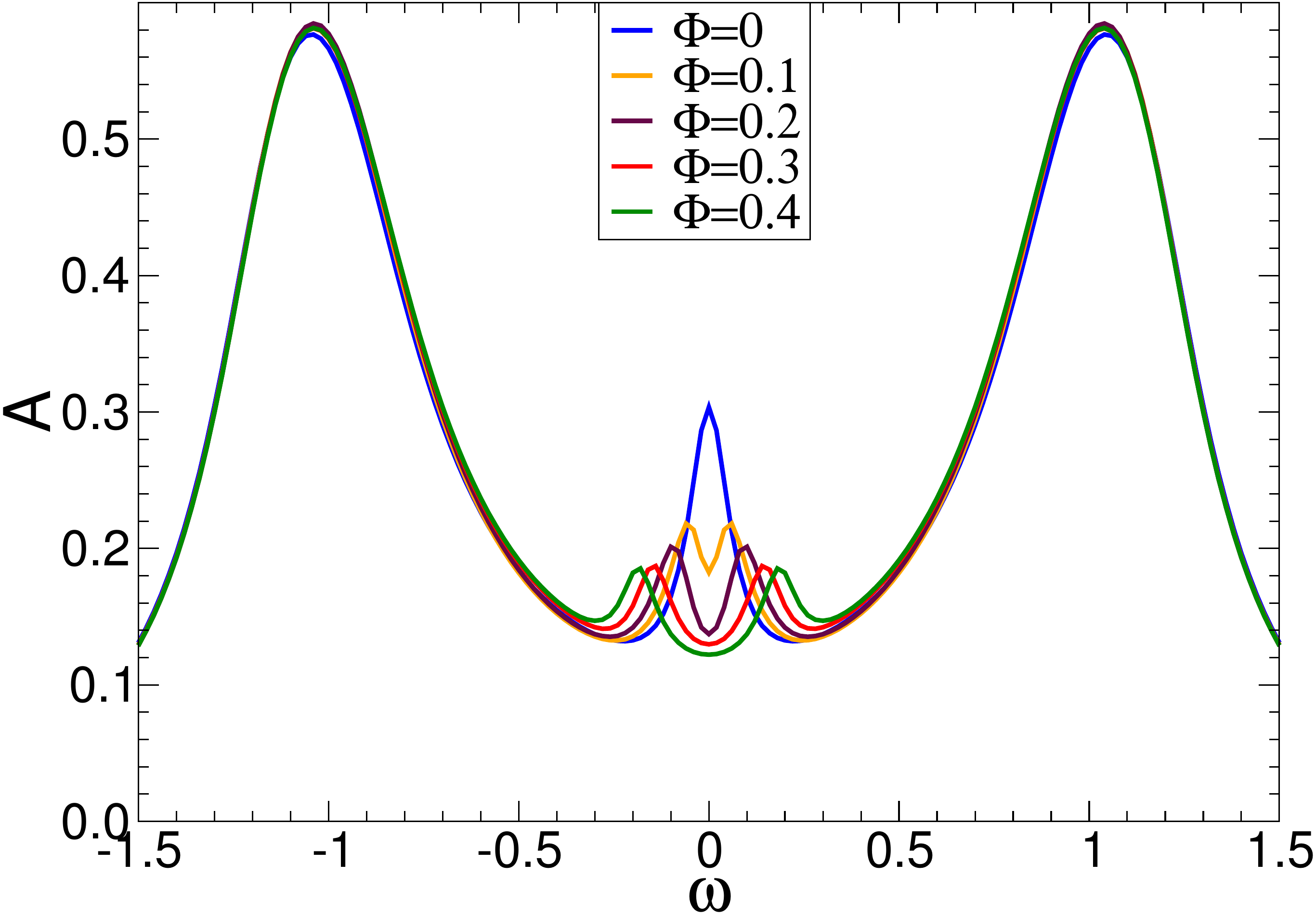}
\caption{(Color online) Single particle spectral function for $N_b = 6$, $U=2$, $t_c=0.1$ and for different values of $\Phi$. Other parameters are as in Fig.~\ref{Green_comp}.}
\label{Resonance_splitting}
\end{figure}

\begin{figure}
\includegraphics[width=0.75\columnwidth]{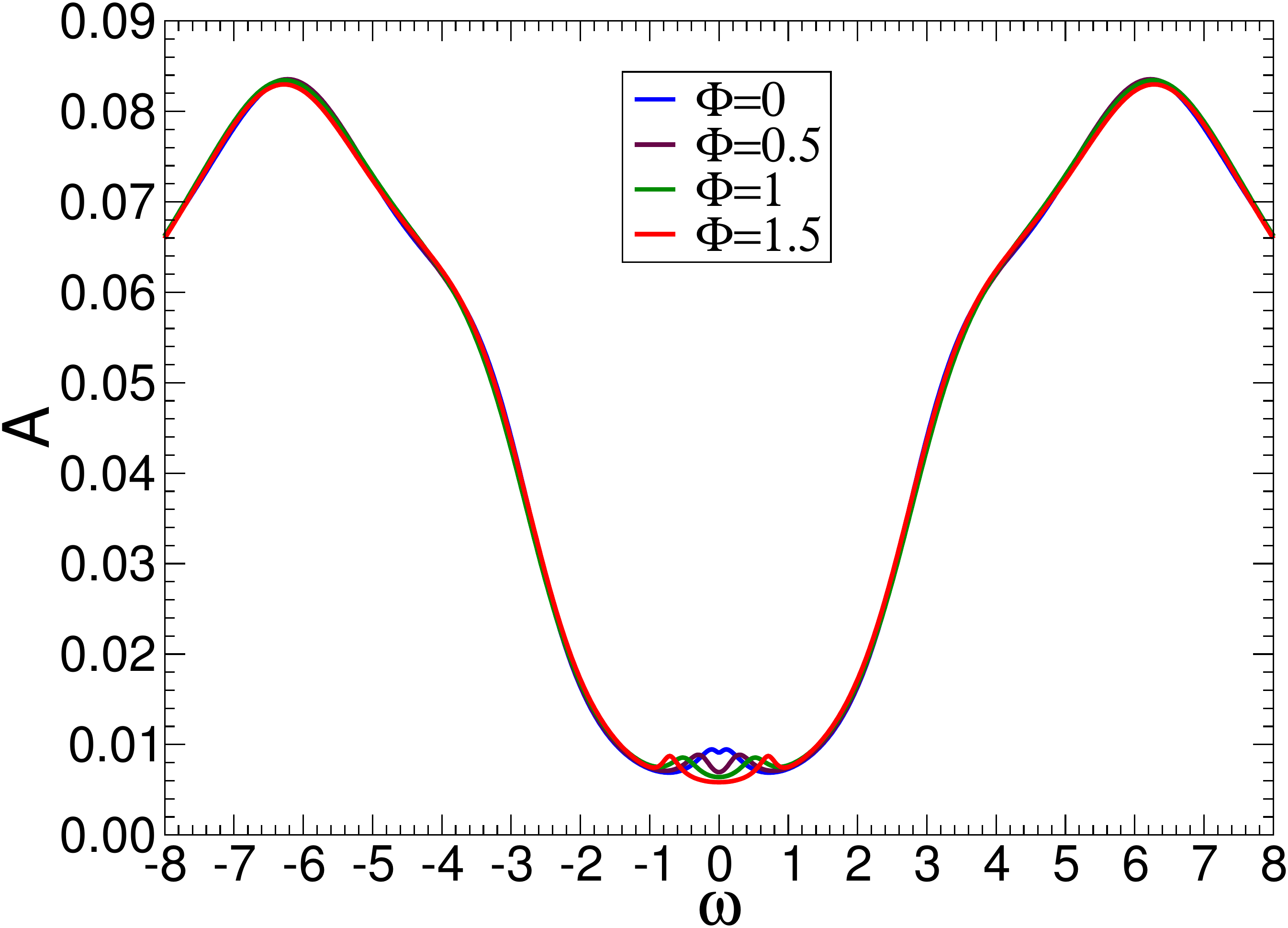}
\caption{(Color online) Single particle spectral function for $N_b = 6$, $U=12$, $t_c=1$ and for different values of $\Phi$. Other parameters are as in Fig.~\ref{Green_comp}.}
\label{Aomega_U12_LBV}
\end{figure}

\begin{center}
\begin{figure*}
\subfigure[]{
\label{Aomega_Phi0}
\begin{minipage}[b]{0.3\textwidth}
\centering \includegraphics[width=1\textwidth]{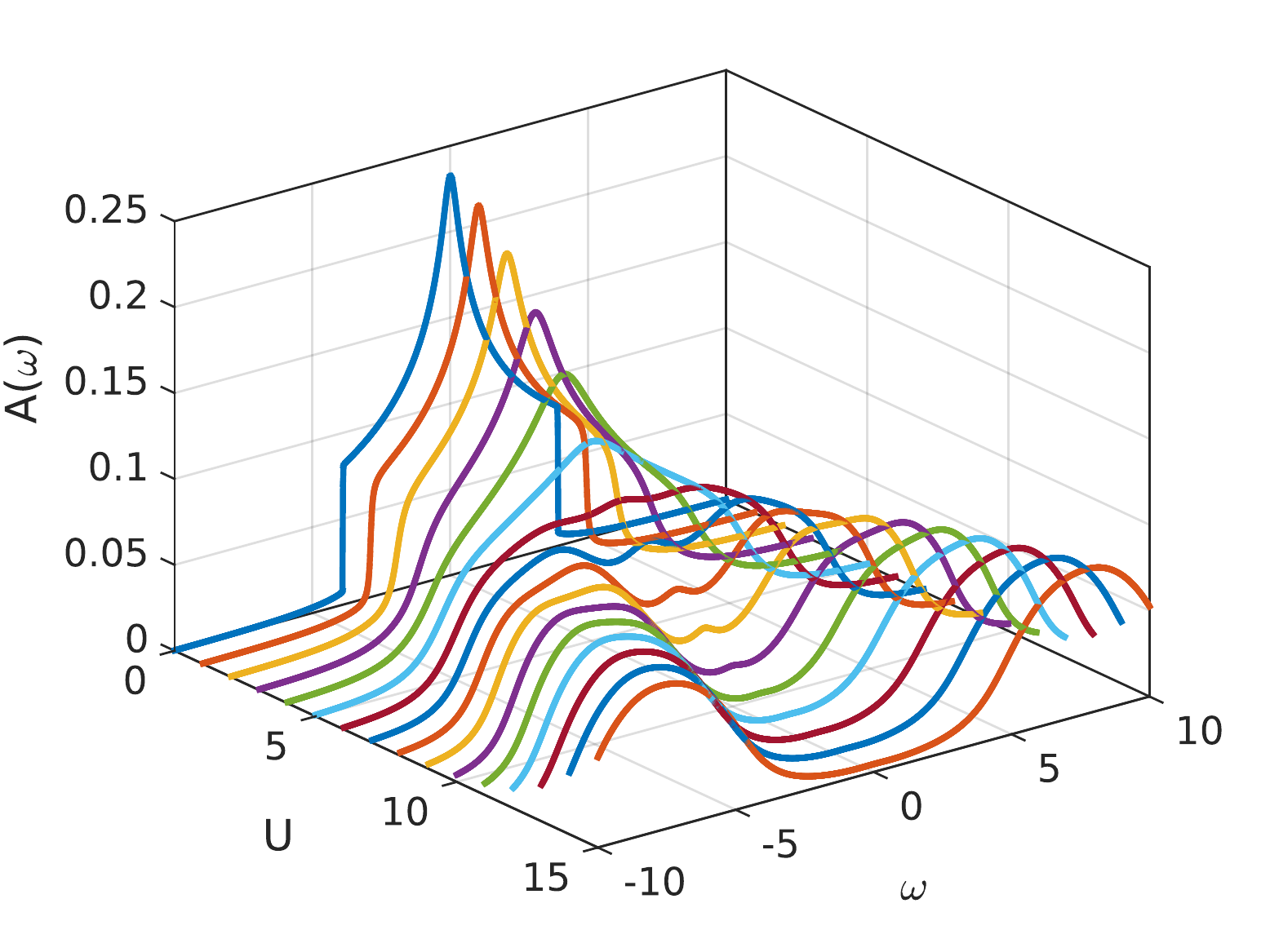}
\end{minipage}}
\subfigure[]{
\label{Aomega_Phi3.5}
\begin{minipage}[b]{0.3\textwidth}
\centering \includegraphics[width=1\textwidth]{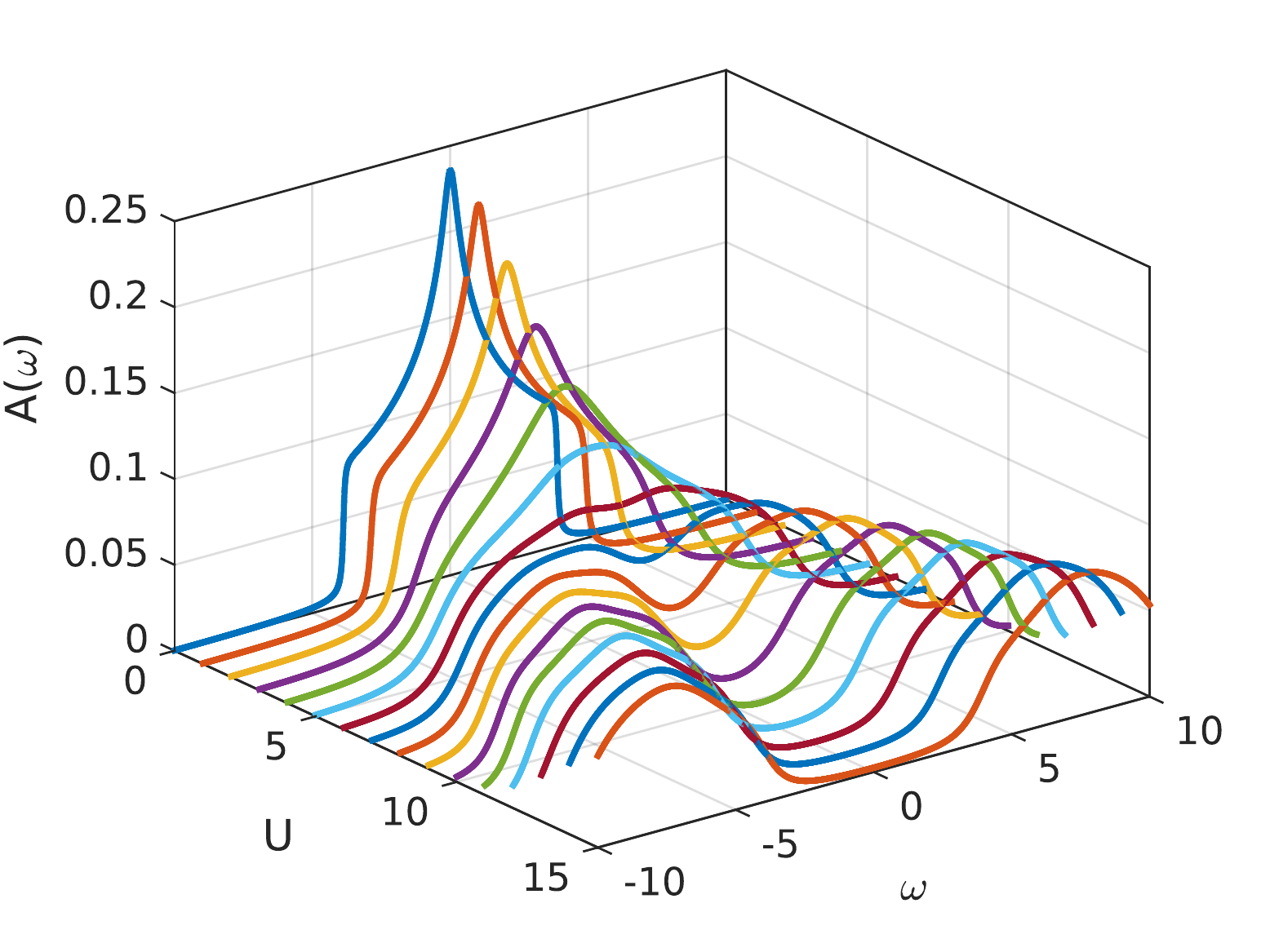}
\end{minipage}}
\subfigure[]{
\label{Aomega_Phi15}
\begin{minipage}[b]{0.3\textwidth}
\centering \includegraphics[width=1\textwidth]{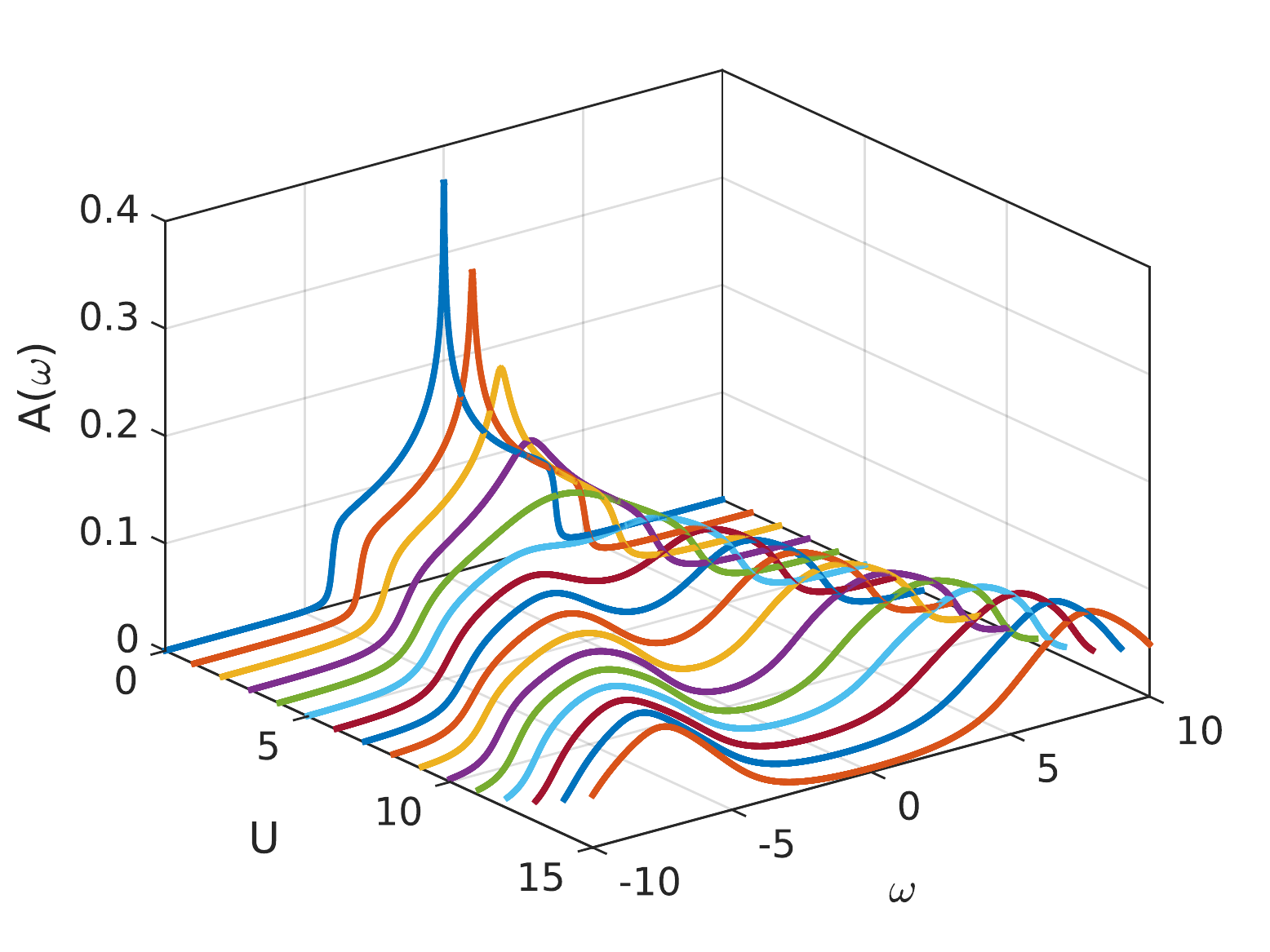}
\end{minipage}}\\
\caption{(Color online) Single particle  spectral function for $N_b = 4$, different values of $U$, $\Phi=0$ (a), $\Phi=3.5$ (b) and $\Phi = 15$ (c). Other parameters are as in Fig.~\ref{Green_comp}.}
\label{Aomega_fix_Phi}
\end{figure*}
\end{center}

\subsection{Non-equilibrium spectral function}\label{spectral_function}
To gain further insight into the properties of the steady state we also investigate the non-equilibrium spectral function, which can be calculated from the Green's function via ${A(\omega)=-\frac{1}{\pi}\Imm G^R(\omega)}$. The results are shown in Fig.~\ref{Aomega_fix_U} for $U=2$ and $U=12$. For weak interaction ($U=2$) the spectral function $A(\omega)$ displays  for all bias voltages a peak at $\omega=0$ and hardly visible Hubbard satellites at the approximate position $\omega =\pm U/2$. The spectral function for $U=2$ (Fig. \ref{Aomega_U2_color1})  depends only very weakly on bias voltage. This is in contrast to the spectral density of the one-dimensional SIAM for which{, it is found that,}   the Kondo peak splits up as a function of voltage and two resonances are observed at the corresponding chemical potentials of the two leads\cite{do.nu.14, wi.me.94, le.sc.01, ro.kr.01,ko.sc.96, fu.ue.03, sh.ro.06, fr.ke.10, nu.he.12, ande.08, ha.he.07, co.gu.14}.  The difference between the two situations is that in the case of the single impurity model the resonance and the Hubbard subbands are clearly separated. In contrast, for the correlated layer and for the set of parameters considered here{, when the isolated correlated layer is metallic}, they overlap due to the broadening induced by the  hopping $t_c$ within the correlated layer.  And indeed if we artificially reduce $t_c$ to $0.1$ (keeping all other parameters fixed) we observe, for $\Phi=0$ a resonance, which is clearly separated from the Hubbard subbands (cf.  Fig.~\ref{Resonance_splitting}). 
In this case,
{
the isolated layer would be insulating and the broadening of the resonance is not any more due to $t_c$, but to an effective energy scale $T_K$, which can be seen as a Kondo temperature. This  originates from a combination of coherent scattering from the leads into the insulating layer, as well as from a self-consistent DMFT process as discussed in Ref.~\onlinecite{he.pe.13}. 
In addition to the broadening mentioned above there is also a spurious broadening due to the limited accuracy of our calculation. Nevertheless, our resolution is sufficient in order to observe a  splitting of the resonance into two peaks at $\mu_{l/r}=\pm V_{B}/2$ as a function of voltage as in  the single impurity case.}

Now we turn to strong interactions ($U=12$), for which the results are depicted in Figs. \ref{Aomega_U12_color1} and \ref{Aomega_U12_LBV}. In equilibrium, i.e. for $\Phi=0$, the hybridization with the leads produces a weak Kondo resonance at the Fermi Energy ($\omega=0$)~\cite{footnote_Two_Peaks}. A nonzero bias voltage splits the resonance into two peaks at $\mu_{l/r}=\pm{\Phi/2}$ (see Fig.~\ref{Aomega_U12_LBV}). For $\Phi \gtrsim3$ the peaks merge into the Hubbard subbands and the spectral function $A(\omega)$ consists of two Hubbard  subbands at the approximate position $\omega =\pm U/2$, while $A(\omega=0)$ is strongly suppressed. For these larger values of $\Phi$ the effect of the bias voltage is small: it modifies only slightly the position and height of the Hubbard subbands. Notice that in order to resolve the Kondo peak and its splitting at low bias we need to use an auxiliary system with $N_b=6$. While this allows to resolve the peaks, the limited accuracy makes them broader than they should be and therefore the Friedel sum rule is not satisfied {\rm even for equilibrium. The reason for this is that a spurious broadening originating from the limitation of our approach reduces the height of the peak at the Fermi level.~\cite{footnote_Total_Weight}} To fulfill this one would have to use more bath sites, and consequently adopt a matrix-product state based solution of the auxiliary system, as we did in Ref.~\onlinecite{do.ga.15} for the Anderson impurity model. This, however, in combination with the DMFT self-consistency, would increase considerably the required computation time.

Next we study the dependence  on the Hubbard interaction $U$ in more detail. Results for equilibrium (${\Phi=0}$), low (${\Phi=3.5}$) and high (${\Phi=15}$) bias voltages are presented in Fig.~\ref{Aomega_fix_Phi}. {As one can see, in the equilibrium case for large $U$ only small excitations are visible at $\omega=0$. Here the Friedel sum rule  would require the $\omega=0$ peak to display  a $U$-independent height.~\cite{footnote_Friedel} However, as discussed above} for large $U$ our approach cannot resolve $T_k$ and the peak becomes strongly suppressed, and for intermediate $U$ the sum rule is not satisfied, as for Fig.~\ref{Aomega_U12_LBV}. Still, Fig.~\ref{Aomega_fix_Phi}(a) displays a crossover from a regime in which the local Fermi liquid peak is already present in the isolated correlated layer (for $U$ below the 2D Mott transition), into a Kondo-Fermi liquid regime in which the peak is produced by coherent spin-flip processes across the Mott insulator, originating from the Fermi levels of the leads. Outside of the Kondo regime, the behavior is qualitatively similar in the three cases. In the non-equilibrium cases, upon increasing the interaction the height of the spectral function at $\omega=0$ decreases and for $U \gtrsim 6$ the spectral function displays a local minimum at $\omega=0$ instead of a maximum, which becomes vanishingly small with increasing $U$. Comparison of Figs.~\ref{Aomega_Phi3.5} and \ref{Aomega_Phi15} shows that for higher bias voltages this resonance disappears at smaller $U$. {Also, for higher bias voltages ($\Phi>10$) the non-interacting spectral function has a sharper peak at $\omega=0$. This is due to the fact that for high bias voltage the leads' density of states do not  overlap any more  and correspondingly states close to the Fermi level do not dissipate any more into the leads, therefore the density of states close to the Fermi level is just the two dimensional density of states, which  features a  logarithmic divergence.} Another effect, clearly visible in Figs.~\ref{Aomega_Phi0},~\ref{Aomega_Phi3.5} and \ref{Aomega_Phi15}, is  the linear shift of the position of the Hubbard subbands with increasing $U$. The peak position is given by $\omega_{\pm} \simeq \pm U/2$.

\section{Conclusions}\label{Conclusions}

{We have presented an improved application of a  DMFT technique for non-equilibrium situations that allows to study directly steady state properties of strongly correlated devices. Like in equilibrium DMFT, the only approximation is the locality of the  self-energy, while the accuracy of the non-equilibrium impurity problem is controlled by the number of bath sites $N_b$ which are attached to Lindblad environments. We find that the accuracy increases exponentially with $N_b$, both  in and out of equilibrium. The approach is benchmarked for a strongly correlated layer coupled to two metallic leads.} While the results in Ref.~\onlinecite{ar.kn.13} for this model were obtained by full diagonalisation of the auxiliary impurity problem and were, thus, restricted to $N_b=2$, here we invoked the non-hermitian Krylov-space method, which allows us to use larger values for $N_b$. 

{With the Krylov-space solver we were able to go up to $N_b=6$. For more bath sides (up to $N_b\gtrapprox 14$) the  MPS solver~\cite{do.ga.15} could be used, but the Krylov-space solver has the advantage to be quicker and more flexible, which is important for the DMFT iteration. For the single layer device studied here we found that $N_{b}=4$ already yields very reliable results in most parameter cases. Only the Kondo regime requires larger values for $N_b$, but with $N_b=6$ at least semi-quantitative results can be achieved.}

We have  investigated the  current-voltage characteristics across a correlated layer. At low bias voltages, we have observed a linear behavior for weak interactions, while the current was exponentially suppressed for strong interactions.~\cite{footnote_Tiny_Current} On the other hand, for higher bias voltages we have observed a reversed picture, whereby the current is larger in the strongly interacting case. In addition we have investigated the current $J$ as a function of the local Hubbard interaction for low as well as for high bias voltages. For lower bias voltages we found that the $J$ decreases monotonically with $U$, while for higher bias voltages, the current  first increases, reaches its maximum and then decreases again. 
The origin of this behavior can be explained by different scattering processes.

In addition to the  current  we have also investigated the steady state spectral function. Our results show that for the set of parameters considered in this manuscript, the spectral function is only weakly dependent on the bias voltage in contrast to the single impurity problem. This is due to the fact that the splitting of the Kondo resonance as a function of $\Phi$ is strongly broadened due to the hopping within the correlated layer. As to be expected, the Hubbard satellites depend almost linearly on $U$, like in the equilibrium case.

\begin{acknowledgments}
We thank Markus Aichhorn, Michael Knap and Martin Nuss for valuable discussions. This work was supported by the Austrian Science Fund (FWF): P24081, P26508, as well as SfB-ViCoM project F04103, and NaWi Graz. The calculations were partly performed on the D-Cluster Graz and on the VSC-3 cluster Vienna. 
\end{acknowledgments}


\begin{thebibliography}{98}
\bibitem [{foo({\natexlab{a}})}]{footnote_DAC}%
  \BibitemOpen
  \href@noop {} {} ({\natexlab{a}}),\ \bibinfo {note} {an approach to deal with
  correlated systems out of equilibrium involving a double analytical
  continuation from Matsubara space has been developed in
  Ref.~\onlinecite{ha.he.07,di.we.10}}\BibitemShut {NoStop}%
\bibitem [{foo({\natexlab{b}})}]{footnote_Convergence}%
  \BibitemOpen
  \href@noop {} {} ({\natexlab{b}}),\ \bibinfo {note} {in practice we compare
  two consecutive hybridization functions in the DMFT iteration and use as
  convergence criterion that the root-mean-square deviation becomes smaller
  than $10^{-5}$.}\BibitemShut {Stop}%
\bibitem [{foo({\natexlab{d}})}]{footnote_Tiny_Current}%
  \BibitemOpen
  \href@noop {} {} ({\natexlab{d}}),\ \bibinfo {note} {there should be a tiny
  contribution to the {\color{dgreen}conductance} (of the order of $T_K$) for
  $\Phi$ much smaller than the insulating gap. However, this is hard to
  resolve. See also the discussion of the Kondo regime below.}\BibitemShut
  {Stop}%
\bibitem [{foo({\natexlab{e}})}]{footnote_Max_Shift}%
  \BibitemOpen
  \href@noop {} {} ({\natexlab{e}}),\ \bibinfo {note} {we expect that with
  increasing bias voltage the maximum of $I=I(U)$ shifts to higher
  interactions.}\BibitemShut {Stop}%
\bibitem [{foo({\natexlab{f}})}]{footnote_Two_Peaks}%
  \BibitemOpen
  \href@noop {} {} ({\natexlab{f}}),\ \bibinfo {note} {if one looks in more
  detail, one can in fact notice two peaks for $\Phi=0$. Within the accuracy of
  our calculation we cannot state whether these are physical or merely an
  artifact.}\BibitemShut {Stop}%
\bibitem [{foo({\natexlab{g}})}]{footnote_Total_Weight}%
  \BibitemOpen
  \href@noop {} {} ({\natexlab{g}}),\ \bibinfo {note} {\color{dgreen}We expect
  the total weight of the peak not to be substantially affected by this
  broadening}\BibitemShut {NoStop}%
\bibitem [{foo({\natexlab{h}})}]{footnote_Friedel}%
  \BibitemOpen
  \href@noop {} {} ({\natexlab{h}}),\ \bibinfo {note} {\color{dgreen}The
  argument is the same as for the Anderson impurity model: Since we are in the
  Fermi liquid phase the imaginary part of the self-energy is zero at
  $\omega=0$. The real part vanishes in the present particle-hole symmetric
  case.}\BibitemShut {Stop}%
\end{thebibliography}
\begin{thebibliography}{98}
\bibitem [{foo({\natexlab{a}})}]{footnote_DAC}%
  \BibitemOpen
  \href@noop {} {}\bibinfo {note} {an approach to deal with
  correlated systems out of equilibrium involving a double analytical
  continuation from Matsubara space has been developed in
  Ref.~\onlinecite{ha.he.07,di.we.10}}\BibitemShut {NoStop}%
\bibitem [{foo({\natexlab{b}})}]{footnote_Convergence}%
  \BibitemOpen
  \href@noop {} {}\bibinfo {note} {in practice we compare
  two consecutive hybridization functions in the DMFT iteration and use as
  convergence criterion that the root-mean-square deviation becomes smaller
  than $10^{-5}$.}\BibitemShut {Stop}%
\bibitem [{foo({\natexlab{d}})}]{footnote_Tiny_Current}%
  \BibitemOpen
  \href@noop {} {}\bibinfo {note} {there should be a tiny
  contribution to the {conductance} (of the order of $T_K$) for
  $\Phi$ much smaller than the insulating gap. However, this is hard to
  resolve. See also the discussion of the Kondo regime below.}\BibitemShut
  {Stop}%
\bibitem [{foo({\natexlab{e}})}]{footnote_Max_Shift}%
  \BibitemOpen
  \href@noop {} {}\bibinfo {note} {we expect that with
  increasing bias voltage the maximum of $I=I(U)$ shifts to higher
  interactions.}\BibitemShut {Stop}%
\bibitem [{foo({\natexlab{f}})}]{footnote_Two_Peaks}%
  \BibitemOpen
  \href@noop {} {} \bibinfo {note} {if one looks in more
  detail, one can in fact notice two peaks for $\Phi=0$. Within the accuracy of
  our calculation we cannot state whether these are physical or merely an
  artifact.}\BibitemShut {Stop}%
\bibitem [{foo({\natexlab{g}})}]{footnote_Total_Weight}%
  \BibitemOpen
  \href@noop {} {} \bibinfo {note} {We expect
  the total weight of the peak not to be substantially affected by this
  broadening}\BibitemShut {NoStop}%
\bibitem [{foo({\natexlab{h}})}]{footnote_Friedel}%
  \BibitemOpen
  \href@noop {} {} \bibinfo {note} {The
  argument is the same as for the Anderson impurity model: Since we are in the
  Fermi liquid phase the imaginary part of the self-energy is zero at
  $\omega=0$. The real part vanishes in the present particle-hole symmetric
  case.}\BibitemShut {Stop}%
\end{thebibliography}

%

\end{document}